\documentclass[conference]{IEEEtran}

\usepackage{cite}
\usepackage{amsmath,amssymb,amsfonts}
\usepackage{amssymb}
\usepackage{algpseudocode}
\usepackage{amsfonts}
\usepackage{graphicx}
\usepackage{fancyhdr}
\usepackage{cases}
\usepackage{textcomp}
\usepackage{extarrows}
\usepackage{algorithm,algpseudocode}
\usepackage{multirow}
\usepackage{graphicx}
\usepackage{overpic}
\usepackage{graphicx}
\usepackage{booktabs}
\usepackage[caption=false,font=footnotesize]{subfig}
\IEEEoverridecommandlockouts

\usepackage{multirow,tabularx}
\usepackage{mathtools}
\usepackage{xcolor}
\usepackage[english]{babel}
\usepackage{amsthm}

\IEEEoverridecommandlockouts
\def\BibTeX{{\rm B\kern-.05em{\sc i\kern-.025em b}\kern-.08em
    T\kern-.1667em\lower.7ex\hbox{E}\kern-.125emX}}
\begin{document}

\bstctlcite{BSTcontrol}
\title{
Energy-efficient Beamforming for RISs-aided Communications: Gradient Based Meta Learning
}

\author{
    \IEEEauthorblockN{Xinquan Wang$^{1,2}$, Fenghao Zhu$^{1}$, Qianyun Zhou$^{1}$, Qihao Yu$^{1}$, Chongwen Huang$^{1,2}$,\\
    Ahmed Alhammadi$^{3}$, Zhaoyang Zhang$^{1}$,  Chau Yuen$^{4}$,~\IEEEmembership{Fellow,~IEEE}, and M\'{e}rouane~Debbah$^{5,}$$^{6}$,~\IEEEmembership{Fellow,~IEEE}
    }
    \IEEEauthorblockA{$^1$ College of Information Science and Electronic Engineering, Zhejiang University, 310027, Hangzhou, China}
    \IEEEauthorblockA{$^2$ State Key Laboratory of Integrated Service Networks, Xidian University, Xi’an 710071, China,}
    \IEEEauthorblockA{$^3$ Technology Innovation Institute, 9639 Masdar City, Abu Dhabi, UAE}
    \IEEEauthorblockA{$^4$ School of Electrical and Electronics Engineering, Nanyang Technological University, Singapore}
    \IEEEauthorblockA{$^5$ KU 6G Research Center, Khalifa University of Science and Technology, P O Box 127788, Abu Dhabi, UAE}
    \IEEEauthorblockA{$^6$ CentraleSupelec, University Paris-Saclay, 91192 Gif-sur-Yvette, France}
}
\maketitle

\vspace{-5mm}
\begin{abstract}
Reconfigurable intelligent surfaces (RISs) have become a promising technology to meet the requirements of energy efficiency and scalability in future six-generation (6G) communications. However, a significant challenge in RISs-aided communications is the joint optimization of active and passive beamforming at base stations (BSs) and RISs respectively.
Specifically, the main difficulty is attributed to the highly non-convex optimization space of beamforming matrices at both BSs and RISs, as well as the diversity and mobility of communication scenarios. To address this, we present a greenly gradient based meta learning beamforming (GMLB) approach.  Unlike traditional deep learning based methods which take channel information directly as input, GMLB feeds the gradient of sum rate into neural networks. Coherently, we design a differential regulator to address the phase shift optimization of RISs. Moreover, we use the meta learning to iteratively optimize the beamforming matrices of BSs and RISs.
These techniques make the proposed method to work well without requiring energy-consuming pre-training. Simulations show that GMLB could achieve higher sum rate than that of typical alternating optimization algorithms with the energy consumption by two orders of magnitude less.
\end{abstract}


\section{Introduction}\label{sec:intro}
\par
The sixth-generation (6G) is expected to integrate sensing and AI-related capabilities into wireless communication systems, and support enriched and immersive experience, enhanced ubiquitous coverage, and enable new forms of collaboration. Reconfigurable intelligent surfaces (RISs) that be deemed to a potential technology for 6G to improve the coverage and spectrum efficiency \cite{6G} has attracted much attention. Specifically, RISs are smart surfaces comprising a large number of passive reflective elements, which can change the incident electromagnetic waves in customized way. The combination of passive phase shifting technique at RISs and active beamforming technique\cite{Beamforming} at base stations (BSs) offers promising solutions for frequency-, energy- and cost-efficiency communications in a variety of scenarios.
However, the joint optimization of beamforming matrices, including the phase shift matrice at RISs and precoding matrice at BSs is a highly non-convex and NP-hard problem\cite{liu2021risconvey}. Therefore, it is a challenge to design the effective beamforming algorithm to take full advantage of RISs to enhance wireless communications.

To address this issue, some novel optimization methods were proposed. Specifically, \cite{OptimalBF} was one of the early attempts to maximize the sum rate problem by a block coordinate descent  based optimization method for the RISs-aided communication systems that renders a stationary solution. Two computationally affordable approaches to solve the non-convex beamforming optimization problem, i.e., the gradient descent and sequential fractional programming that were proposed in \cite{HCW2019TWC}, while \cite{Pero2021AO} proposed projected gradient algorithm for RISs-aided communications and used a Lipschitz constant to guarantee the convergence.
However, the most of these methods involve the calculating inversions of large scale matrices, which is a cubic complexity operation that results in very heavy computational overhead especially in the millimeter wave communication scenario.

\par On the other hand, some researchers resort to the artificial intelligence based method since it is proved to perform great in various scenarios. Specifically, \cite{huang2020ddpg,zhu2023robust,chen2023ddpg} proposed the deep reinforcement learning method with deep deterministic policy gradient to ensure that action and state spaces are continuous. However, this class of algorithms involve the intensive computational overhead, and be lack of the robustness as well. To address the robust issue, meta learning based methods such as \cite{di2023mcddpg,JYXia,jung2021meta} were proposed to cooperate the optimization of all the beamforming matrices. However, researchers at the university of Massachusetts show the energy consumption for training the
neural networks (NNs) are huge. In other words, the training process can emit more than 626,000 pounds of carbon dioxide equivalent--nearly five times the lifetime emissions of the average car \cite{AIpower}.
\par To solve above-mentioned issues, we propose the gradient based meta learning beamforming (GMLB) method with the extremely low energy consumption to solve the joint optimization problem in RISs-aided communications. The main features of GMLB include:


\begin{itemize}
    \item \emph{Free of Pre-training:} Our proposed method is no need for the pre-training, which significantly reduce the training energy consumption, and ensures its robustness across scenarios as well.
    \item \emph{Gradient as Input:} Instead of directly feeding channel matrices into NNs, GMLB computes the gradients of the inputs with respect to the sum rate, and then feeds these gradients into NNs.
    \item \emph{Differential Regulation:} We treat the outputs of the NNs as differentials of the beamforming matrices. To tackle the issue of the triangular function cycle in the optimization of RIS elements, GMLB uses a custom differential regulator to regulate the outputs of the NN.
    \item \emph{Small-scale Neural Networks:} The gradient-as-input mechanism allows GMLB to use small-scale NNs, which further reduces the energy consumption.
\end{itemize}

\par
The rest of this paper is organized as follows.
The system model and the problem formulation will be described in Section \ref{sec:sys}.
A comprehensive analysis of the GMLB method, including explanation of the NNs' architecture the principles behind the differential regulator is provided in Section \ref{sec:GMLB}.
Simulation results are provided in Section \ref{sec:simulation} to verify the performance of the proposed algorithms, and the conclusions are presented in Section \ref{sec:conclusion}.


\section{System Model and Problem Formulation}\label{sec:sys}
\begin{figure}[t]
	\begin{center}
		\centerline{\includegraphics[width=0.85\linewidth]{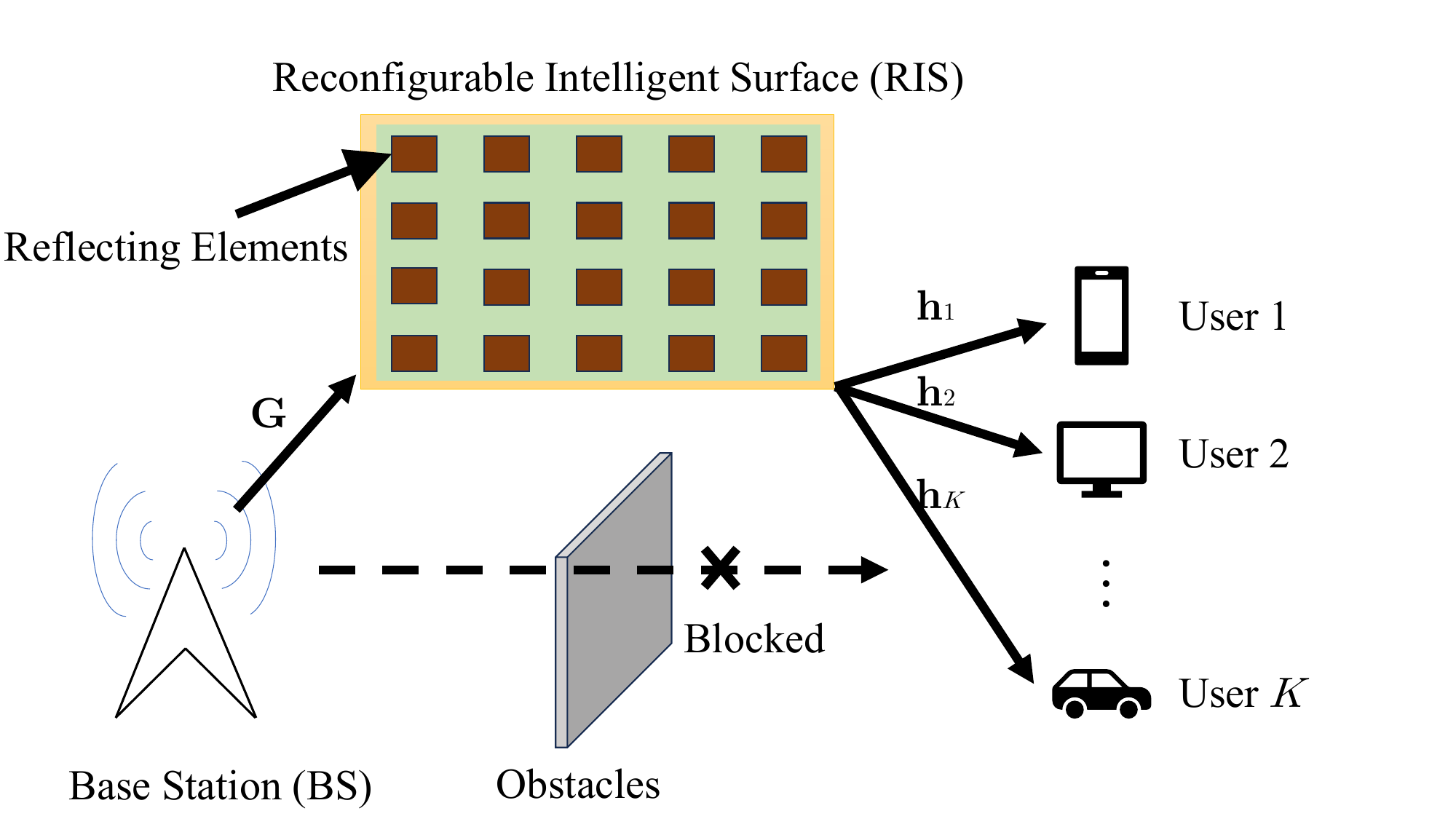}}\vspace{-3mm}
		\caption{RISs-aided MU-MISO beamforming system.}
		\label{fig:scenario} \vspace{-8mm}
	\end{center}
\end{figure}
\addtolength{\topmargin}{0.03in}
\par Consider a multi-user multiple-input-single-output (MU-MISO) communication system with a $M$-antenna BS and a RIS with $N$ reflect elements which serves $K$ single antenna users, as shown in Fig. \ref{fig:scenario}. $K$ data streams with unit power marked as  $\mathbf{s}\in \mathbb{C}^{K\times 1}$ satisfying $E\left\{\mathbf{s}\mathbf{s}^H\right\}=\mathbf I$ are transmitted simultaneously, each of its elements targeting at one of the $K$ users. We assume that the direct links between the BS and the users are blocked, and all the channels, including the channel $\mathbf{G} \in \mathbb{C}^{N \times M}$ between BS and RIS and the channel $\mathbf{H} \in \mathbb C^{K\times N}$
between RIS and users are perfectly known. The signal received at the $k$-th user is given as
\begin{equation}\label{kth User Received}
	y_k=\mathbf{h}^H_k\mathbf{\Theta G w}_k s_k+\sum_{i\neq k}^K \mathbf{h}^H_k \mathbf{\Theta G w}_i s_i + n_k,
\end{equation}
where $y_k$ denotes the signal received by the $k$-th user, and $s_k$ is the $k$-th element of $s$. The phase shift matrix is defined as
$\mathbf \Theta=\text{diag}[e^{j\theta_1}, e^{j\theta_2}, \cdots, e^{j\theta_N}]$, in which $\theta_n $ represents the phase shift of $n$-th RIS element. $\mathbf{w}_k$ is the $k$-th column of the precoding matrix $\mathbf W\in \mathbb{C}^{M\times K}$ of the BS, and $\mathbf{h}_k$ is the transpose of the $k$-th row of $\mathbf{H}$. $n_k$ is the the additive circular white Gaussian noise with zero mean and variance $\sigma^2$ of the $k$-th user.

\par To maintain the power constraint of the BS, a constraint for $\mathbf{W}$ is proposed as
\vspace{-2mm}\begin{equation}\label{Power Constraint}\vspace{-0mm}
	\mathrm{Tr}(\mathbf{W}^ H\mathbf{W})\leq P,
\end{equation}
where $P$ is the total transmitted power of the BS; $(\cdot)^{{H}}$ denotes the hermitian transpose, and $\mathrm{Tr}(\cdot)$ represents the trace of matrix. In \eqref{kth User Received}, the $\mathbf{h}^H_k\mathbf{\Theta G w}_k s_k$ is the desired
signal at the $k$-th user and $\sum_{i\neq k}^K \mathbf{h}^H_i \mathbf{\Theta G w}_i s_i$ is treated
as interference between users.
Since $s_k$ is unit-powered, the signal to interference plus noise ratio (SINR) of the $k$-th user could be denoted as
\begin{equation}\label{snr single user}
	\gamma_k = \frac{||\mathbf{h}_k^H\mathbf{\Theta G w}_k||^2}{\sigma^2 + \sum_{j \neq k}^K||\mathbf{h}_k^H\mathbf{\Theta G w}_j||^2},
\end{equation}
To evaluate the performance of systems, the spectrum efficiency is used as a metric, which could be expressed as
\begin{equation}\label{spectrum efficiency}
	R(\mathbf{W,\Theta;H,G})=\sum_{k=1}^K \log_2(1+\gamma_k),
\end{equation}
and the loss function be denoted as $L=-R$.

\begin{figure}[t]\vspace{-2mm}

    \centering
    \hfill
    \label{fig:convex}
    \subfloat[Highly non-convex rank-0 space.]{%
      \includegraphics[width=0.32\textwidth]{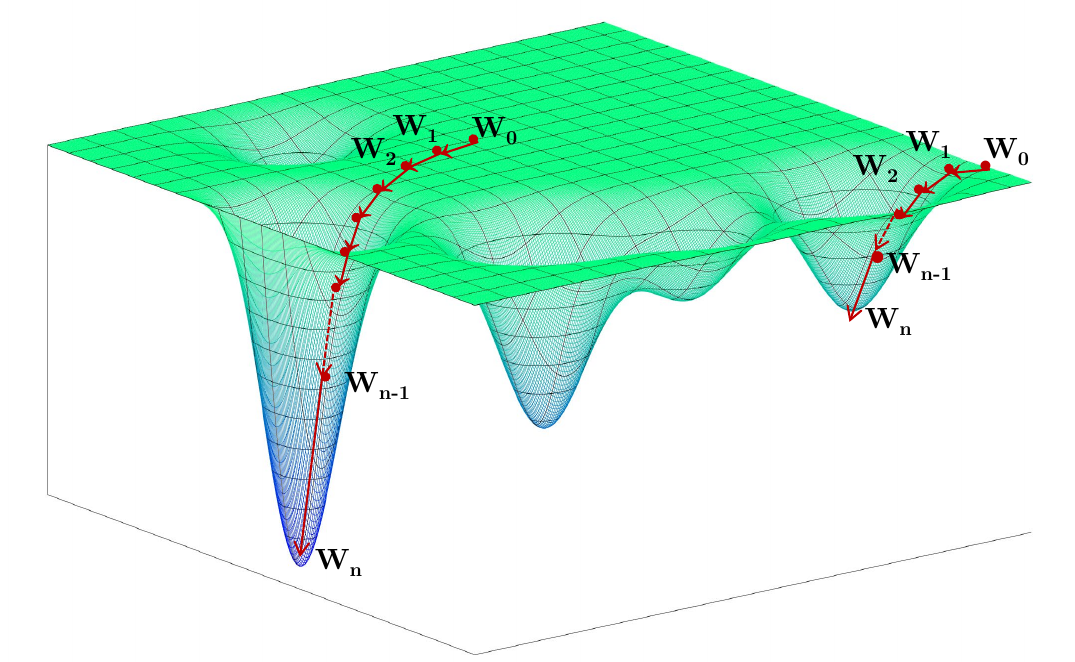}
      \label{fig:convex1}
    }\hfill
    \vspace{-0mm}
    \subfloat[Smoother rank-1 gradient space with less local minima.]{%
      \includegraphics[width=0.42\textwidth]{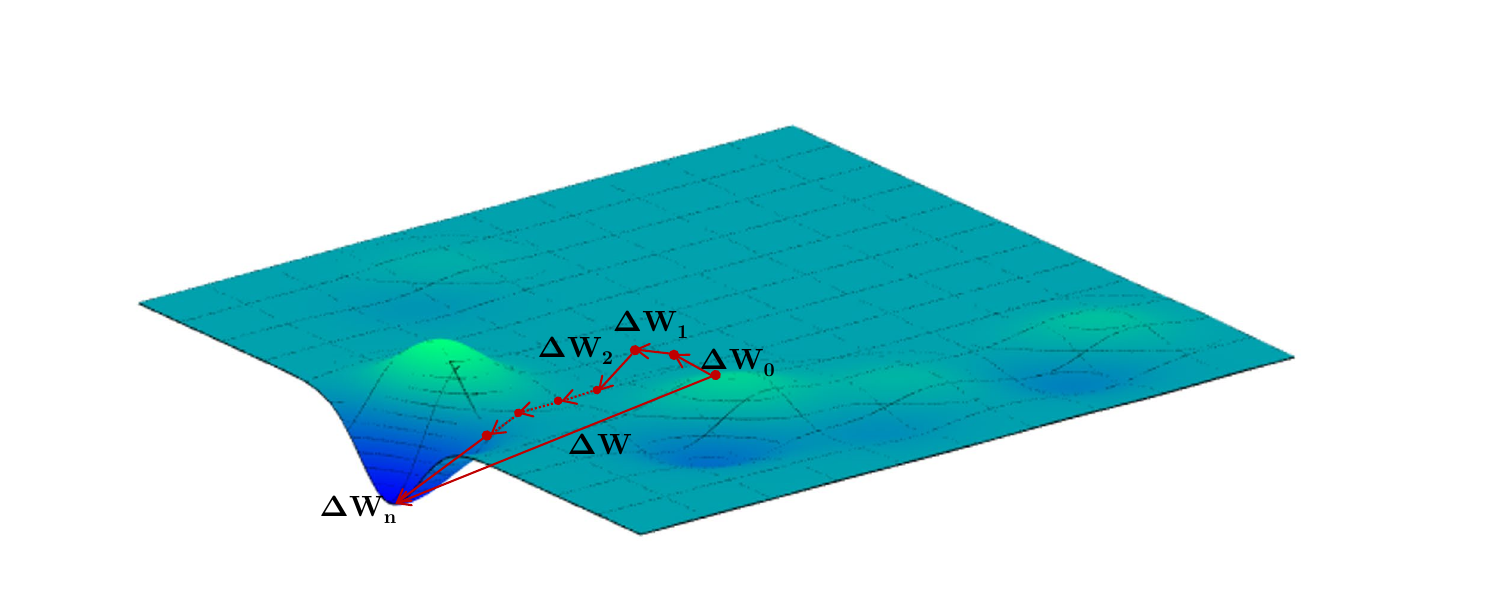}
      \label{fig:convex2}
    }
    \caption{Optimization landscapes for RISs-aided communications.}
    \vspace{-4mm} 
\end{figure}

\par The aim of this paper is to increase the sum rate $R$ described in \eqref{spectrum efficiency} through jointly optimizing  $\mathbf \Theta$ and $\mathbf W$ in any given scenario. This optimization problem could be formulated as
\begin{equation}\label{optimization problem}
    \begin{split}
        \mathop{\rm {max}}\limits_{\substack{\mathbf{W}\in \mathcal{W}\; \\ \mathbf{\Theta}\in\mathcal{O}}} & R(\mathbf{W,\Theta;H,G}),\\
	     \mathrm{s.t.\ \ }&\mathrm{Tr}(\mathbf{W}^{H}\mathbf{W})\leq P,\\
	    &\mathbf{\Theta}=diag[e^{j\theta_1},e^{j\theta_2},\cdots,e^{j\theta_N}],\\
	    &|\theta_j|=1,j=1,2,3,\cdots,N.
    \end{split}
\end{equation}
Here $\mathcal{W}$ and $\mathcal{O}$ are feasible regions for $\mathbf{W}$ and $\mathbf{\Theta}$, respectively.

\section{Gradient Based Meta learning Beamforming}\label{sec:GMLB}
To address the non-convex optimization problem \eqref{optimization problem}, we propose an unsupervised GMLB that is given in this section.
\subsection{Proposed GMLB Algorithm}\label{subsec:algorithm}
\par Traditional beamforming scenarios without RISs are generally characterized by direct channel models and present a smoother optimization landscape. Prior solutions have obtained optimized precoding matrices by searching directly in the feasible region (rank-0 space).
However, the introduction of RISs and the associated constraints makes the joint  beamforming optimization problem more difficult, and also leads to easily coverage the potential local minima and saddle points in the rank-0 space, where $\nabla R(\mathbf{W,\Theta;H,G})=\mathbf{0}$ as shown in Fig. \ref{fig:convex1}.
\par However, when we focus the gradient of the sum rate with respect of $\mathbf{W}$ and $\mathbf{\Theta}$ over the entire rank-0 space, we find that after taking differentiation, some of the areas that were previously considered as the local minima no longer show the characteristics of extreme points.
Therefore, we propose an GMLB approach by taking differentiation to address the phase shift optimization of RISs. Instead of directly feeding raw channel matrices into NNs, we first compute the gradients of the sum rate with respect to initialized $\mathbf{W}_0$ and $\mathbf{\Theta}_0$,
and subsequently feed these gradients into NNs.
Accordingly, the outputs of the NNs are treated as differentials of $\mathbf{W}_0$ and $\mathbf{\Theta}_0$, which are denoted as $\mathbf{\Delta W},\mathbf{\Delta \Theta}$, respectively.
Therefore, the highly non-convex optimization problem \eqref{optimization problem} can be optimized by searching in a smoother landscape (rank-1 space), as illustrated in Fig. \ref{fig:convex2}. We denote the feasible regions as $\nabla \mathcal{W},\nabla \mathcal{O}$ for $\mathbf{\Delta W},\mathbf{\Delta \Theta}$, respectively, and the transformed optimization problem could be expressed as
\vspace{-0mm}\begin{equation}\label{New Optimization Problem}
    \begin{split}
         \mathop{\rm {max}}\limits_{\substack{\Delta\mathbf{W}\in \nabla \mathcal{W}\\ \Delta\mathbf{\Theta}\in\nabla \mathcal{O}}} & R(\mathbf{W}^*,\mathbf{\Theta}^*;\mathbf{H,G}),\\
	    \mathrm{s.t.\quad}& \mathrm{Tr}[(\mathbf{W}^*)^{H}\mathbf{W}^*]\leq P,\\
	    &\mathbf{\Theta^*}=diag[e^{j\theta_1^*},e^{j\theta_2^*},\cdots,e^{j\theta_N^*}],\\
	    &|\theta_j|=1,j=1,2,3,\cdots,N.
    \end{split}
\end{equation}
Here $\mathbf{W^*=W}_0+\Delta\mathbf{W}$ and $\mathbf{\Theta^*=\Theta}_0\cdot(\Delta\mathbf{\Theta})$ are the updated beamforming matrices.



\begin{figure}[t]\vspace{-2mm}
	\begin{center}
		\centerline{\includegraphics[width=0.95\linewidth]{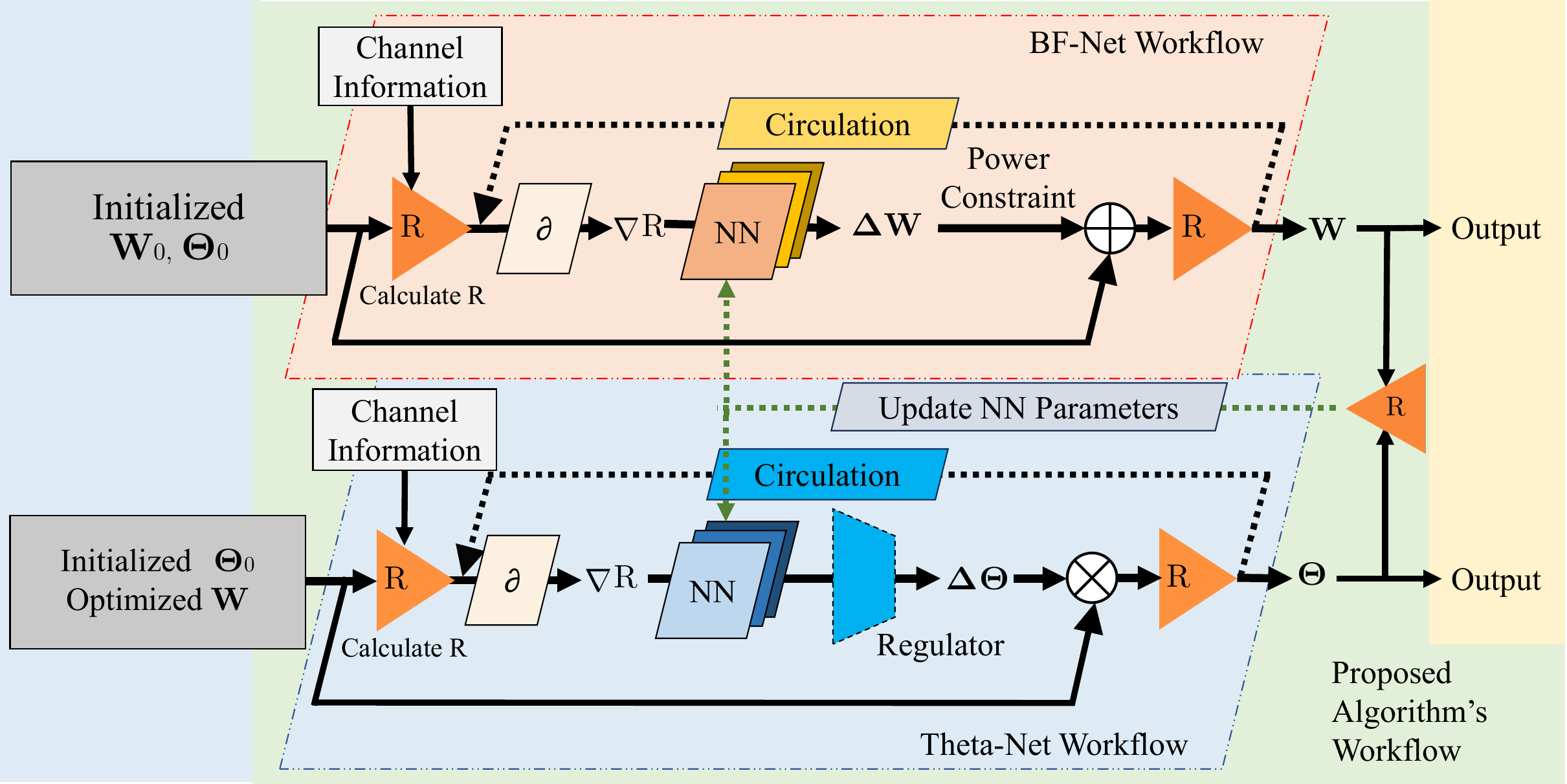}}\vspace{-2mm}
		\caption{Workflow of implemented GMLB.}
		\label{fig:GMLB}\vspace{-10mm}
	\end{center}
\end{figure}
\subsection{The Network Architecture}
\addtolength{\topmargin}{0.06in}


\par In this paper,
we apply a meta learning framework to gradient-as-input mechanism
as illustrated in Fig. \ref{fig:GMLB}. Specifically, we propose a model that integrates two NNs: the \emph{BF-Net} to optimize $\mathbf{W}$, and the \emph{Theta-Net} to optimize $\mathbf{\Theta}$. Note that the NNs are designed to be small-scale for energy-efficiency, as detailed in the subsequent subsections.
After the forward propagation of \emph{BF-Net} and \emph{Theta-Net}, GMLB computes the loss function by $L_i = -R(\mathbf{W}_i,{\mathbf{\Theta}_i};\mathbf{ H, G})$ and perform backpropagation to compute the gradients of the NN parameters $\mathbf{\theta_W,\theta_\Theta}$, which are denoted as $\nabla_{\mathbf{\theta_W}}$ and $\nabla_{\mathbf{\theta_\Theta}}$, respectively. Then the parameters are updated by
\begin{equation}\label{t1}
    \mathbf{\theta_W}^*=\mathbf{\theta_W}-\alpha_{\mathbf{W}}\cdot \mathrm{Adam}(\nabla_{\mathbf{\theta_W}},\theta_{\mathbf{W}}),
\end{equation}
\begin{equation}\label{t2}
    \mathbf{\theta_\Theta}^*=\mathbf{\theta_\Theta}-\alpha_{\mathbf{\Theta}}\cdot \mathrm{Adam}(\nabla_{\mathbf{\theta_\mathbf{\Theta}}},\theta_{\mathbf{\Theta}}),
\end{equation}
where the $\mathbf{\theta_W}^*$ and $\mathbf{\theta_\Theta}^*$ are updated NN parameters, and $\alpha_{\mathbf{W}}$ and $\alpha_{\mathbf{\Theta}}$ are learning rates for \emph{BF-Net} and \emph{Theta-Net}, respectively.
Completed algorithm is outlined in Algorithm \ref{alg:meta}.


\subsection{BF-Net Design}\label{BF-Net}
\par Given the channel information, the relationship between $\mathbf{W}^*$ and $R$ is described as
\begin{equation}\label{rank2 H grad}\vspace{1mm}
        R_{\mathbf{W}^*}=\sum_{k=1}^K \log_2(1+ \frac{||\mathbf{h}_k^H\mathbf{\hat\Theta G} (\mathbf{w}_k+\mathbf{\Delta w}_k)||^2}{\sigma^2 + \sum_{j \neq k}^K||\mathbf{h}_k^H\mathbf{\hat\Theta G} (\mathbf{w}_j+\mathbf{\Delta w}_j)||^2}),
\end{equation}
where $\hat{\mathbf{\Theta}}$ represents the $\mathbf \Theta$ acquired from either initialization or previous steps, and $\mathbf{\Delta w}_{k}$ denotes the $k$-th column of $\mathbf{\Delta W}$.

\begin{algorithm}[t]
\caption{GMLB Workflow }
\label{alg:meta}
\begin{algorithmic}[1]
\Procedure{GMLB}{$\mathbf{H}$, $\mathbf{G}$}
    \State Initialize $\mathbf{\theta_W,\theta_\Theta},\mathbf{W}_0,\mathbf{\Theta}_0$.
    \For{$i\leftarrow 1,2,\cdots,N_e$}
        \State Compute sum rate $R_0 = R(\mathbf{W}_{i-1},{\mathbf{\Theta}}_{i-1};\mathbf{ H, G})$;
        \State $\mathbf{W}^{(0)}_i\leftarrow \mathbf{W}_{i-1}$;
        \For{$j\leftarrow 1,2,\cdots,N_r$}
            \State Calculate gradient $\mathbf \nabla R(\mathbf{W}^{(j-1)}_i,{\mathbf{\Theta}}_{i-1};\mathbf{ H, G})$;
            \State Calculate $\Delta \mathbf{W}^{(j)}_i$ with \emph{BF-Net}($\mathbf \nabla R;\theta_\mathbf{W}$);
            \State $\mathbf{W}^{(j)}_i\leftarrow \mathbf{W}^{(j-1)}_i+\Delta \mathbf{W}^{(j)}_i$;
            \State Regulate the power of BS antennas as \eqref{Power Constraint};
            \State Compute $R = R(\mathbf{W}^{(j)}_i,{\mathbf{\Theta}}_{i-1};\mathbf{ H, G})$;
            \If{$R < R_0$} Revert to previous values;
            \Else $\ R_0\leftarrow R$.
            \EndIf

        \EndFor
        \State $\mathbf{W}_i\leftarrow\mathbf{W}^{(N_r)}_i$, $\mathbf{\Theta}^{(0)}_i\leftarrow\mathbf{\Theta}_{i-1}$;
        \State Compute sum rate $R_0 = R(\mathbf{\Theta}_{i-1},\mathbf{W}_i;\mathbf{ H, G})$;
            \For{$j\leftarrow 1,2,\cdots,N_r$}
                \State Calculate gradient $\mathbf \nabla R(\mathbf{\Theta}^{(j-1)}_i,\mathbf{W}_i;\mathbf{ H, G})$;
                \State Calculate $\Delta \mathbf{\Theta}^{(j)}_i$ with \emph{Theta-Net}($\mathbf \nabla R;\theta_\mathbf{\Theta}$);
                \State $\mathbf{\Theta}^{(j)}_i\leftarrow \mathbf{\Theta}^{(j-1)}_i\cdot\Delta \mathbf{\mathbf{\Theta}}^{(j)}_i$;
                \State Process $\mathbf{\Delta \Theta}^{(j)}_i$ with differential regulator;
                \State Compute $R = R(\mathbf{\Theta}^{(j)}_i,{\mathbf{W}}_i;\mathbf{ H, G})$;
                \If{$R < R_0$} Revert to previous values;
                \Else $\ R_0\leftarrow R$.
                \EndIf
            \EndFor
        \State $\mathbf{\Theta}_i\leftarrow\mathbf{\Theta}^{(N_r)}_i$;
        \State Compute loss function $L_i = -R(\mathbf{W}_i,{\mathbf{\Theta}_i};\mathbf{ H, G})$;
        \State Backpropagate $L_i$;
        \State Update NN parameters $\mathbf{\theta_W,\theta_\Theta}$ by \eqref{t1} and \eqref{t2}.
    \EndFor
    \State \Return Optimized $\mathbf{W}_{N_e}$, $\mathbf{\Theta}_{N_e}$.
\EndProcedure
\end{algorithmic}
\end{algorithm}
\begin{table}[t] \vspace{-4mm}
    \centering
    \caption{Number of Neurons in the NNs}\vspace{-2mm}
    \label{BFNN}
    \begin{tabular}{c c c c}
     \toprule
     No.&Layer Name &\emph{BF-Net}&\emph{Theta-Net}  \\
     \midrule
     1&Input Layer&$2\times K$&$N$\\
     2&Linear Layer 1 & $200$   &$200$\\
     3&ReLU Layer & $200$ &$200$ \\
     4&Linear Layer 2 & $ 2\times K$ &$N$\\
     5&Differential Regulator Layer&/&$N$\\ [1.0ex]
     \bottomrule
    \end{tabular}\vspace{-5mm}
\end{table}
\par Equation \eqref{rank2 H grad} displays a more linear relationship than \eqref{rank2 theta grad} because $\mathbf{w}_k$ is simply pre-multiplied by $\mathbf{h}_k\mathbf{\Theta G}$, and the elements of $\mathbf{W}$ itself are all linear.
By utilizing the GMLB algorithm detailed in Algorithm \ref{alg:meta}, a NN with a few layers can effectively capture the characteristics of \eqref{rank2 H grad}. The details of the proposed \emph{BF-Net} can be found in Table \ref{BFNN}.

\vspace{-1mm}\subsection{Theta-Net Design}
\par Similar to the analysis for \emph{BF-Net}, we could notify the association between $\mathbf \Theta^*$ and the $R$ as
\begin{equation}\label{rank2 theta grad}\vspace{1mm}
        R_{\mathbf{\Theta}^*}=\sum_{k=1}^K \log_2(1+ \frac{||\mathbf{h}_k^H(\mathbf{\Theta}\cdot \mathbf{\Delta \Theta})\mathbf{G}\mathbf{\hat w}_k||^2}{\sigma^2 + \sum_{j \neq k}^K||\mathbf{h}_k^H(\mathbf{\Theta}\cdot \mathbf{\Delta \Theta})\mathbf{G} \mathbf{\hat w}_j||^2}),
\end{equation}
where $\hat{\mathbf{w}}_{(\cdot)}$ represents the $\mathbf {w}_{(\cdot)}$ derived from either initialization or previous steps. Here we represent phase shift vector of RIS in $rad$ with $\overline\theta\in \mathbb{R}^{N\times 1}$. Thus the relation between $\mathbf{\Theta}$ and $\overline{\theta}$ could be denoted as $\mathbf{\Theta}=e^{j\overline{\theta}}$.
Although the $R_{\mathbf{W}^*}$ and $R_{\mathbf{\Theta}^*}$ have similar forms, we find that simply transplanting \emph{BF-Net} to optimize $\mathbf{\Theta}$ does not work.
\par When comparing \eqref{rank2 theta grad} with \eqref{rank2 H grad}, it is evident that optimizing $\mathbf \Theta$ presents multifaceted challenges:
\begin{itemize}\label{item:reason of theta}
    \item
    To calculate the sum rate with \eqref{rank2 theta grad}, relationship of $\mathbf{\Theta^*}=diag[e^{j\mathbf (\overline\theta+\Delta \overline\theta)}]=diag[\cos \mathbf (\overline\theta+\Delta \overline\theta)+j\sin \mathbf (\overline\theta+\Delta \overline\theta)]$ is used.
    In this function, $\mathbf{\Theta}^*$ undergoes periodic variation within the interval $2\pi$. Given its periodic nature, this would cause complications in the optimization if the elements of $(\overline\theta+\Delta\overline \theta)$ exceed the region of monotonic increase of the triangular functions.

    \item
    The matrix $\mathbf{\Theta}$ is positioned between $\mathbf{h}_{(\cdot)}$ and $\mathbf{G}$ in \eqref{snr single user}. Therefore, minor changes to $\mathbf{\Theta}$ may significantly affect the sum rate due to multiplicative interactions, which makes the optimization more unstable and challenging.
\end{itemize}
\par Having identified the challenges in optimizing $\mathbf{\Theta}$, it becomes imperative to address these concerns during the architectural design of our neural network. The cornerstone of our strategy lies in constraining the amplitude of the differential, $\mathbf{\Delta \Theta}$, to ensure stability during optimizations.
\par To this end, we present \emph{Theta-Net} detailed in Table \ref{BFNN}, in which we designed a differential regulator that serves as an inherent element to regulate the outputs, as
\begin{equation}\label{theta-period constraint}
	\mathbf{\Delta \Theta}=\lambda \cdot \sigma(\mathbf{\Delta \Gamma}),
\end{equation}
where $\mathbf{\Delta \Gamma}$ is the output of the fourth layer in Table \ref{BFNN}, $\lambda \cdot \sigma(\cdot)={\lambda}/(1+e^{-(\cdot)})$ is Sigmoid function multiplied by an amplification operator.
This design ensures that the $\mathbf{\Theta}$ is within a limited range, thereby addressing these above-mentioned potential problems.


\section{Simulation Results}\label{sec:simulation}
In this section, we evaluate the performance of the proposed method. The channel matrices $\mathbf{G,h}$ are modeled based on the Rician channel model for two individual links, expressed as
\begin{equation}\label{channel_h}
    \mathbf{h}_{k}=\sqrt{\frac{\kappa_{k}}{1+\kappa_{k}}}\mathbf{h}_{k}^{LoS}
    +\sqrt{\frac{1}{1+\kappa_{k}}}\mathbf{h}_{k}^{NLoS},
\end{equation}
\begin{equation}\label{channel_G}
    \mathbf{G}=\sqrt{\frac{\kappa_G}{1+\kappa_G}}\mathbf{G}^{LoS}
    +\sqrt{\frac{1}{1+\kappa_G}}\mathbf{G}^{NLoS},
\end{equation}
where $\kappa_k$ and $\kappa_G$ are Rician factors of $\mathbf{h}_k$ and $\mathbf{G}$, respectively. $\mathbf{h}_k^{LoS}$ and $\mathbf{G}^{LoS}$ are the line of sight (LoS) components of the channel matrices, while $\mathbf{h}_k^{NLoS}$ and $\mathbf{G}^{NLoS}$ are the none line of sight (NLoS) components of the channel matrices. We set $\kappa_k=\kappa_G=10$ for all generated scenarios and average the results over 50 independent scenarios.


\subsection{Baselines}\label{subsec:benchmark}

\par
\begin{figure}[t]
\centering
    \includegraphics[width=0.8\linewidth]{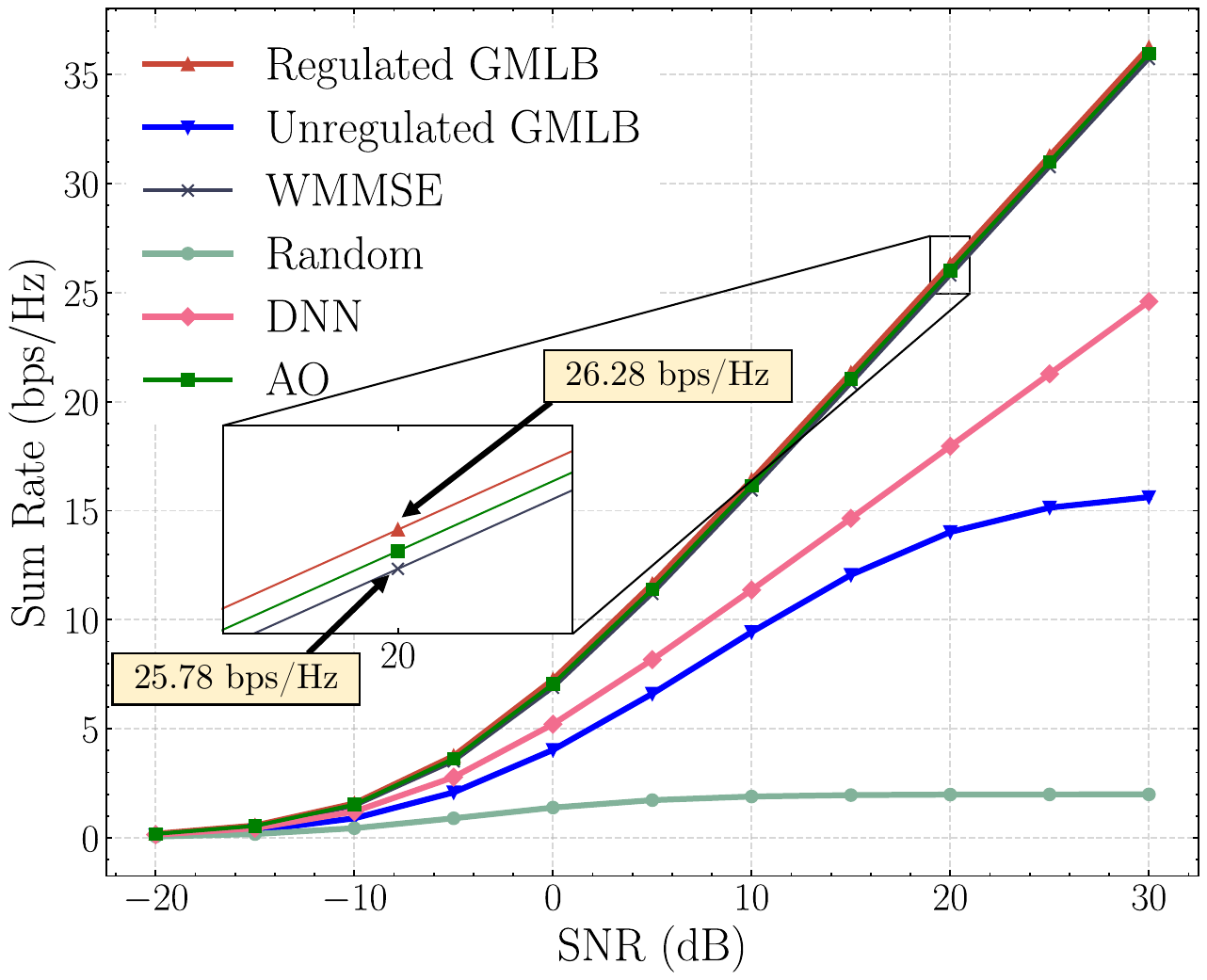}\vspace{-4mm}
    \caption{Performance of the proposed algorithm against the established baselines across varying SNRs.}
    \label{fig:snr}\vspace{-4mm}
\end{figure}

\begin{figure}[t]
    \centering\includegraphics[width=0.81\linewidth]{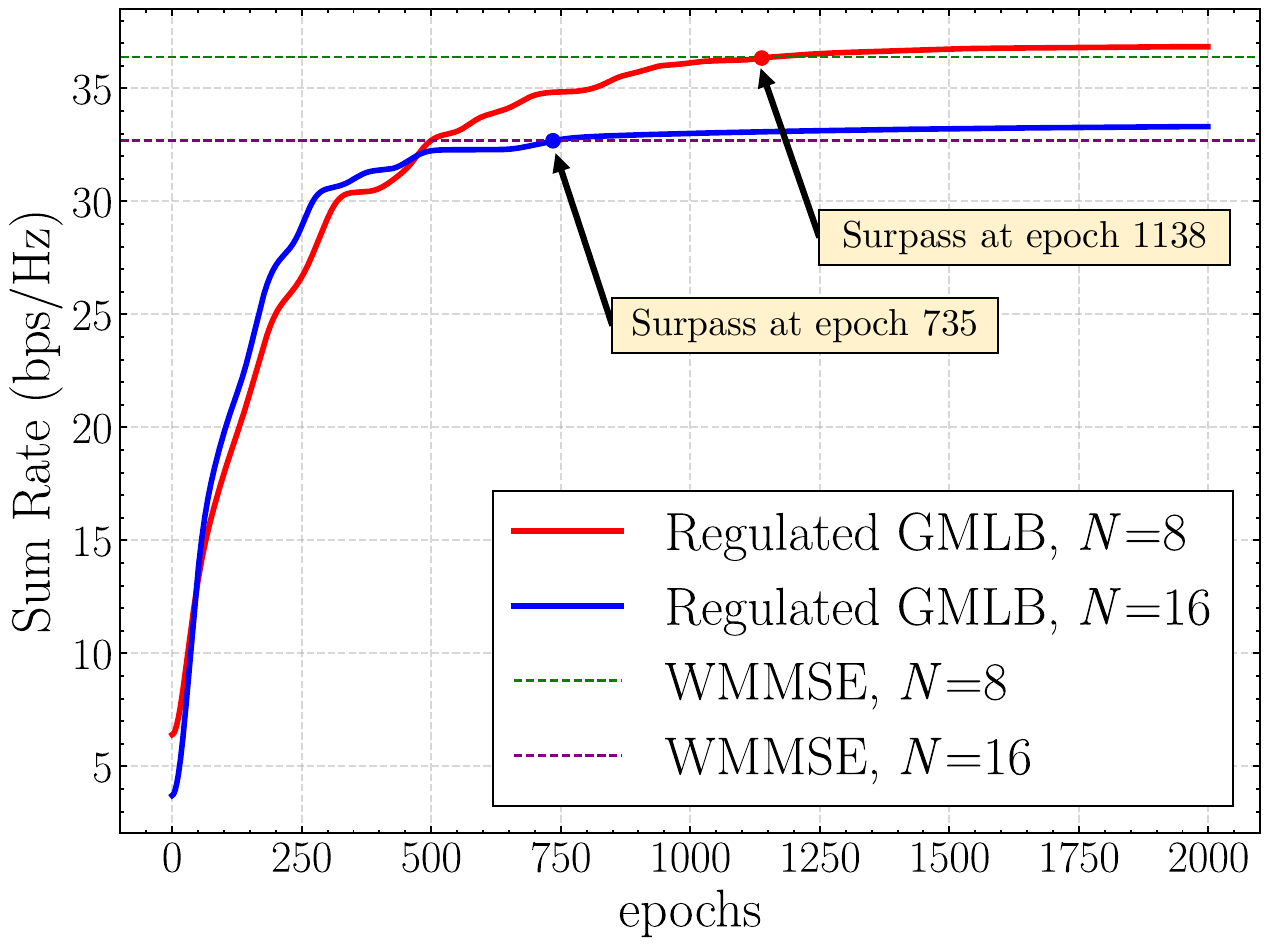}\vspace{-4mm}
    \caption{Comparison of the proposed algorithm's sum rate with the WMMSE reference over epochs.}
    \label{fig:iter_result}\vspace{-4mm} 
\end{figure}

\begin{figure}[t]
    \centering
    \includegraphics[width=0.80\linewidth]{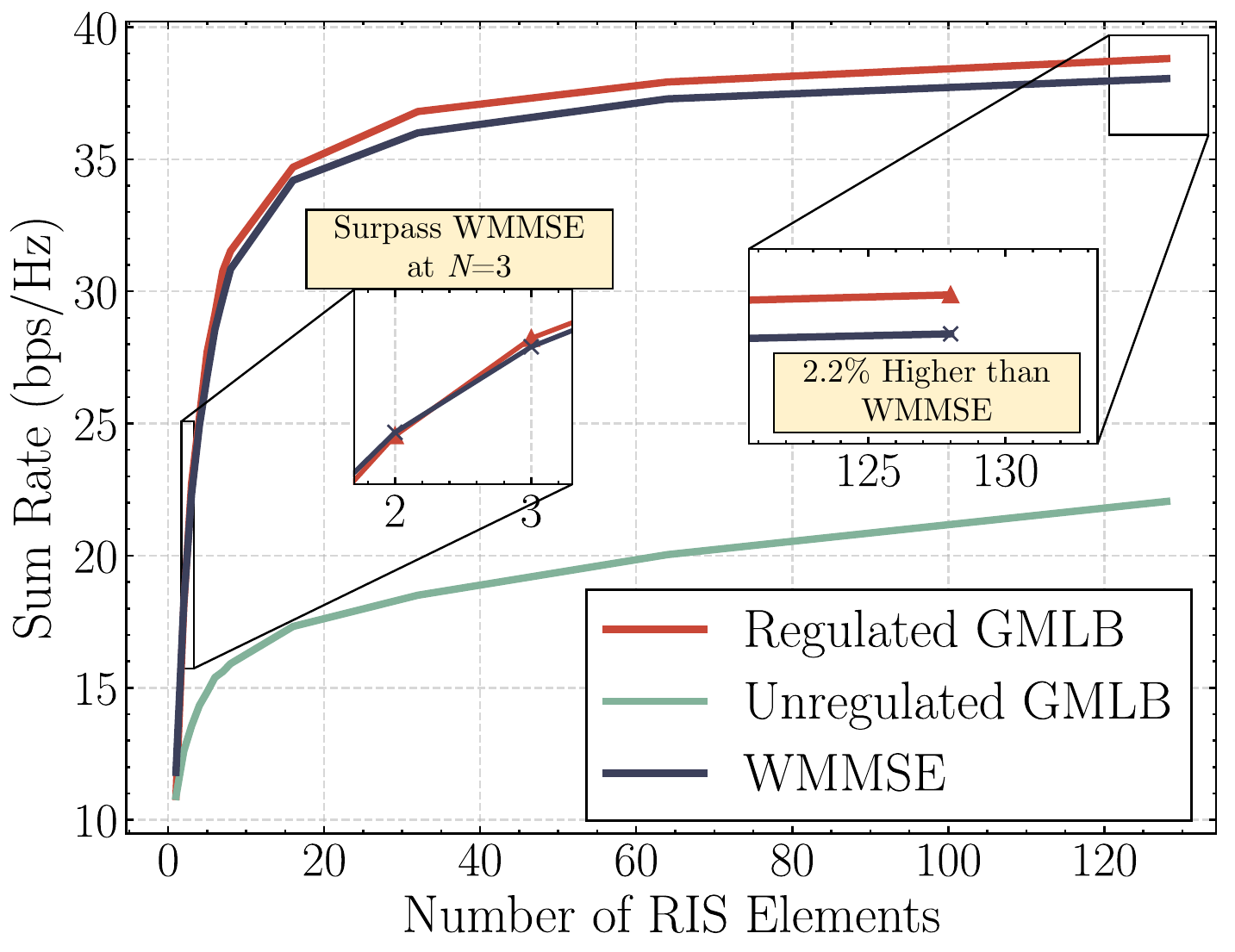}\vspace{-4mm}
    \caption{Performance of the proposed algorithm against the established baselines across varying numbers of RIS elements.}
    \label{fig:ris}\vspace{-4mm}
\end{figure}

 In the simulations, we compare our results with several algorithms. We set $P,N_e,N_r,\alpha_{\mathbf{W}},\mathbf{\alpha_{\Theta}}$ as $1000,5000,1,1\times 10^{-3},1.5\times10^{-3}$, respectively.
We run the simulations on an EPYC 75F3 CPU and a RTX 3090 GPU.
We label the results of the proposed method as \emph{Regulated-GMLB} in Fig. \ref{fig:snr}-\ref{fig:cmp}.
We use the alternating optimization (\emph{AO}), deep neural networks(\emph{DNN}) and weighted minimum mean square error (\emph{WMMSE}) from \cite{quant,WMMSE,OptimalBF}, respectively.
We also include a result named \emph{Random}, which is based on the randomly generated $\mathbf{W}$ and $\mathbf{\Theta}$.
In Fig. \ref{fig:snr} and \ref{fig:ris}, we include a version of GMLB without differential regulation, labeled as \emph{Unregulated-GMLB}, to show the effectiveness of the differential regulator.



\subsection{Impact of SNRs}
We provide a simulation focusing on the performance of GMLB against varying signal-to-noise ratios (SNRs).
Here SNR in dB is defined by the formula $10\log_{10}(P/{\sigma^2})$.
Both the proposed algorithm and the baselines are performed with SNRs ranging from -20dB to +30dB and $M=N=K=4$.
In Fig. \ref{fig:snr}, we observe that the proposed GMLB method outperforms \emph{DNN}, \emph{Random} and \emph{Unregulated-GMLB} significantly, and surpasses \emph{WMMSE} and \emph{AO} slightly. Note that GMLB outperforms \emph{WMMSE} by 1.9\% when SNR is 20dB.

\subsection{Impact of Training Epochs}
\par We present the sum rate over the training process, where we set the SNR to 20dB and set two sets of scenarios where $M,K$ are 4, and $N$ is 4 and 8 respectively.
Fig. \ref{fig:iter_result} displays the simulation results, with the curves smoothed with a window of 5 for clarity. The result shows that sum rates rise rapidly at the beginning, achieving a sum rate relatively close to \emph{WMMSE} quickly. Then, the sum rates continue to rise and surpass the \emph{WMMSE} within a few epochs.
When we compare the two given curves, we find that the rise of computational cost when the scale is enlarged is acceptable, which is a 55\% increase comparing to a doubled number of RIS elements.

\subsection{Impact of Number of RIS Elements}
We evaluated the GMLB based on different numbers of reflecting elements in the RIS. Fig. \ref{fig:ris} shows the results with parameters $M$ and $K$ both set to 4 and a SNR of 20dB.
Both the proposed algorithm and baselines are performed with $N$ varying from 1 to 128.  It can be seen that  the sum rates increase as $N$ increases.
Note that the GMLB method slightly outperforms \emph{WMMSE}, as the domination becomes more apparent as the number of RIS elements increases, as much as 2.2\%.
We find that \emph{Unregulated GMLB} shows its limitation to work in large-scale RIS scenarios, which proves that the introduction of differential regulator enhances GMLB's scalable of large scale RISs significantly.

\subsection{Analysis on Energy Consumption of the Algorithms}
\begin{figure}[h]
    \centering
    \includegraphics[width=0.83\linewidth]{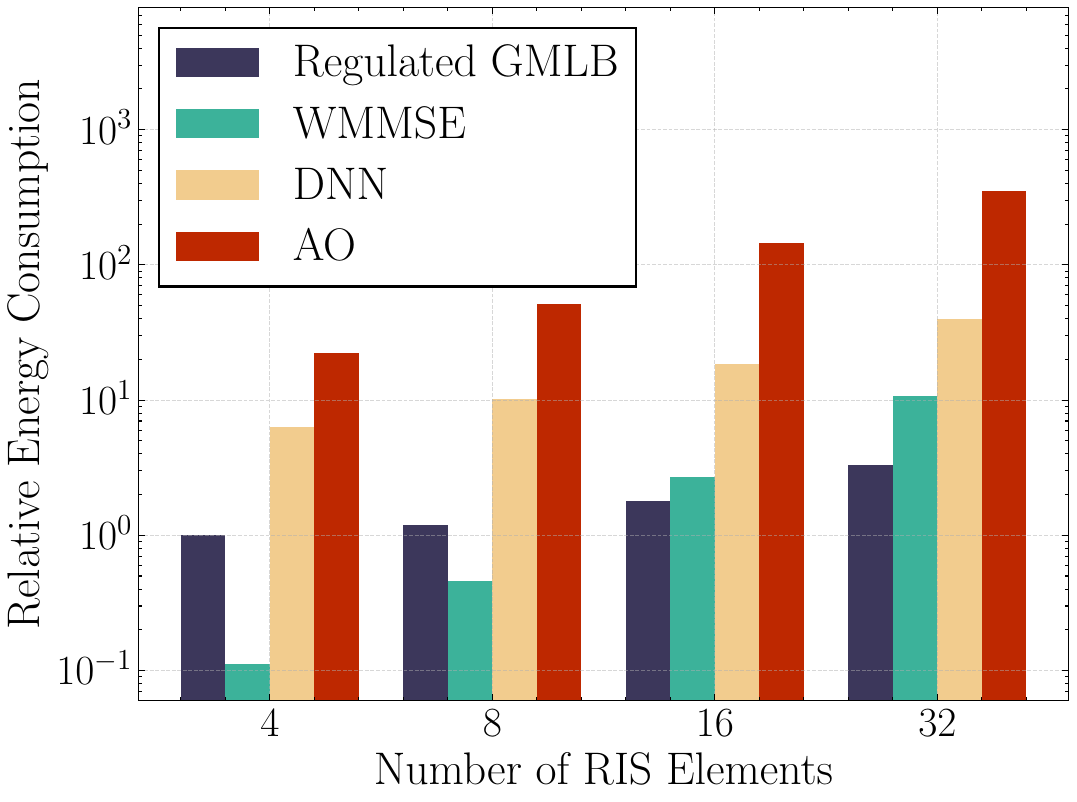}\vspace{-4mm}
    \caption{Relative power consumption of the proposed algorithm against the established baselines across varying numbers of RIS elements,  where GMLB at $N=1$ as a unit.}
    \label{fig:cmp}\vspace{-4mm}
\end{figure}

We present the energy consumption of the proposed GMLB compared with the baselines, where we set the SNR as 20dB and set 4 sets of scenarios where $M,K$ are 4 and $N$ is 4, 8, 16, 32, respectively.
We analyse the energy consumption of the algorithm by observing power consumption of our simulation devices, and normalize the result (GMLB in $N=4$ as unit) for clarity. Fig. \ref{fig:cmp} shows that the energy consumption of the proposed algorithm is similar to that of \emph{WMMSE} when $N$ is very small, but the advantage of energy saving becomes significant as $N$ increases, reducing nearly 70\% energy consumption at $N=32$. It is noticed that the energy consumption of GMLB is far less than that of \emph{DNN} and \emph{AO}, close to reduced energy by two orders of magnitude.
\section{Conclusion}\label{sec:conclusion}
\par
In this paper, a pre-training free method for RISs-aided communications named GMLB was introduced. The features of GMLB enable it to achieve the higher sum rate in a variety of scenarios with less energy-consuming with respect to typical methods.
Specifically, the key feature of GMLB is to use the gradient-as-input mechanism.
Instead of simply feeding channel matrices into NNs, GMLB feeds the gradient of initialized beamforming matrices with respect to the sum rate into NNs.
Furthermore, a differential regulator was designed to address the specific challenges in RISs-aided communications scenarios. Simulations showed that the GMLB outperforms typical baselines.
In the future, we will focus on exploring RISs-aided communications with incomplete channel information and applying our gradient based technique to dynamic scenarios.

\bibliographystyle{IEEEtran}
\bibliography{string}
\vspace{12pt}

\end{document}